\pdfoutput=1
\documentclass[runningheads]{llncs}
\usepackage[T1]{fontenc}
\usepackage{lmodern}   %
\usepackage{amsmath,amssymb}
\usepackage{graphicx}
\usepackage{booktabs,array,tabularx,longtable}
\usepackage{placeins}
\usepackage{xcolor}
\usepackage{url}
\usepackage[hidelinks]{hyperref}
\graphicspath{{figures/}}

\newcommand{\ind}{\mathbf{1}}
\newcolumntype{Y}{>{\raggedright\arraybackslash}X}
\begin{document}
\title{Finite-Key Readiness and Authentication-Key Replenishment in QKD-Assisted UAV Edge Computing}
\titlerunning{Finite-Key Readiness in QKD-Assisted UAV Edge Computing}
\author{Dang H. Le\inst{3}\orcidID{0009-0002-3462-9002} \and
Ban Q. Tran\inst{1,2}\orcidID{0000-0002-4100-5217} \and
Susan Mengel\inst{1}\orcidID{0000-0003-0722-1941}}
\authorrunning{D. H. Le et al.}
\institute{Department of Computer Science, Texas Tech University, TX 79416 USA\\
\email{\{bantran,susan.mengel\}@ttu.edu} \and
Department of Computing Fundamentals, FPT University, Hanoi, 10000, Vietnam\\
\email{bantq3@fe.edu.vn} \and
SAP Innovation Lab, FPT University, Hanoi, 10000, Vietnam\\
\email{dangisden@gmail.com}}
\maketitle
\begin{abstract}
A service that needs fresh quantum keys must wait for key release, and may be denied it, before protected computation starts. We quantify this cost for a trusted stationary UAV serving two ground gateways, each obtaining its keys from one finite-key decoy-state BB84 attempt with parameters fixed in advance. A count-level Monte Carlo model, reproduced by a semi-analytic model, links key acceptance to protected upload, computation, and download, including failed attempts and hover energy. At 26~dB effective optical loss, with authentication hashing at its proxy-benchmarked, near-zero cost, fresh keys raise the 95th-percentile completion time of both tasks from 10.37 to 28.65~s; hover energy grows in proportion. The scalar rate models evaluated here miss the deadline limit on the block size. The pulse budget lies in a window bounded below by authentication-key replenishment and above by the deadline; for a fixed source, channel, and acceptance model, the authentication debit shifts the floor, whereas hashing cost does not. At a 35-s deadline the largest tested loss with a non-empty window is 27~dB at 1\% misalignment and 22~dB at 2\%; at 3\% no tested setting has one. Conditional on the assumed acquisition probability, the selected configurations meet this deadline with simultaneous lower confidence bounds of 96.9--97.3\% on held-out trials, at 26~dB and at 18--20~dB with 2\% misalignment. A high-cost stress reference that charges hashing in full narrows the window and lowers these loss limits to 26 and 20~dB at the reference processing capacity. A small rise in error rate or loss pushes the deadline probability well below its target and drains authentication keys, while the protocol's security parameter is unchanged.
\end{abstract}
\section{Introduction}
Quantum key distribution (QKD) establishes a shared key whose secrecy rests on physics rather than on computational hardness, using a quantum channel and an authenticated classical channel~\cite{scarani}. A positive predicted key length does not say \emph{when}, or whether, the key is released: a finite-key protocol fixes the block, tests it, and either outputs a key of predetermined length or aborts~\cite{a11}. For a service that must protect data before computing, release time and abort probability become part of its latency.

Edge computing on an unmanned aerial vehicle (UAV) makes this concrete. A UAV newly deployed over a site, reassigned between clusters, or out of keys from earlier sessions must wait for a QKD attempt before each task. We adopt this fresh-key policy deliberately: a QKD key is shared only with the station at the other end of the optical link, and a stock of application keys for many stations enlarges what a captured aerial node exposes; while operating, the UAV is still trusted. Waiting is costly, because hovering draws 1.6--1.7~kW against tens of watts for the radios and QKD payload~\cite{B8_Drone-_and_Vehicle-Based_Quantum_Key_Distribution}. Since drone QKD now produces finite-key secret keys with ground and vehicle stations~\cite{tian,B8_Drone-_and_Vehicle-Based_Quantum_Key_Distribution}, the cost of placing such keys on the critical path of a computing service is a practical question.

A post-quantum key encapsulation mechanism such as ML-KEM~\cite{D9_Module-Lattice-Based_Key-Encapsulation_Mechanism_Standard} completes in one message exchange but rests on a computational hardness assumption: a recorded exchange can be attacked later if that assumption falls. A finite-key QKD protocol is information-theoretically secure under the assumptions of its proof~\cite{a11}, after which the data rest on AES-256. We quantify the price of this advantage in time, energy, and robustness, and compare the time with an ML-KEM estimate. QKD also consumes pre-shared authentication keys with every attempt, since each post-processing message must be authenticated; this couples the security layer to the service schedule.

We study a trusted stationary UAV serving two ground gateways in sequence, each with one aggregate task and one QKD attempt, at 26-dB effective optical loss, and ask which pulse budgets meet key availability, deadlines, and authentication-key replenishment together, what bounds them, and how an active adversary erodes them.

Our contributions are:
\begin{enumerate}
\item \textbf{A failure-aware service model.} We connect fixed-length finite-key BB84 attempts to protected computation, retaining abort time and hover energy. A semi-analytic model reproduces the simulated 95th completion percentile within 0.03~s, checking approximations rather than physical calibration.
\item \textbf{A deadline--replenishment window.} Authentication-key balance imposes a sufficient lower bound and deadline envelopes an upper bound on the pulse budget. Gross scalar key rates miss both; net rates recover the balance floor but miss the ceiling. We characterize the effects of authentication debit, hashing, loss, and misalignment, and confirm selected configurations on 5,000 held-out pairs with simultaneous bounds.
\item \textbf{Availability under disturbance.} Small error or loss disturbances preserve the inherited security parameter while defeating deadlines and draining authentication keys. Wider margins improve robustness at a quantified latency cost.
\end{enumerate}

\subsection{Related work}
Finite-key decoy-state analyses~\cite{lim,rusca} capture the fluctuations that asymptotic rates omit; we adopt the fixed-length acceptance framework of Wiesemann et al.~\cite{a11}. Satellite finite-block studies~\cite{sidhu,bae} address limited passes; free-space studies~\cite{scriminich} model propagation and coupling, here replaced by fixed loss. Drone QKD has been demonstrated with a ground station~\cite{tian} and with vehicles, at error rates of 2.5--3.4\%~\cite{B8_Drone-_and_Vehicle-Based_Quantum_Key_Distribution}; earlier airborne experiments are reviewed in~\cite{xue2021}. These works report key rates; none places the key on the critical path of a deadline-constrained task.

UAV-enabled mobile edge computing (MEC) optimizes communication and computation~\cite{li2019,li2020,sun}; physical-layer security protects UAV-MEC offloading through the secrecy rate against a passive eavesdropper, without keys~\cite{nguyen2024pls}. UAV-assisted key provisioning~\cite{kong}, QKD-network resource allocation with a key-delivery-delay constraint~\cite{stan}, QKD-secured cloud--edge collaboration~\cite{zhu}, and QuHE~\cite{quhe}, which couples QKD with homomorphic encryption and MEC allocation, treat keys as a rate or utility; we track individual finite-key attempts, their acceptance randomness and failure time, and the resulting deadlines.

QKD is known to be vulnerable to denial of service through noise above the security threshold or exhaustion of authentication keys~\cite{alleaume2014,price2017}, and key-pool depletion has been demonstrated on deployed nodes~\cite{burdiak2023}. These analyses are qualitative or use the asymptotic threshold, for example 11\% for one-way BB84~\cite{alleaume2014}; we quantify how much smaller the tolerable disturbance becomes for finite-key configurations sized for low latency, and its cost in deadlines and authentication keys. Key recycling across attempts~\cite{portmann} could lower the per-attempt cost; we do not assume it.

We do not propose a new security proof or a globally optimal scheduler; Sec.~\ref{sec:limits} states the limitations.
\section{System Model}
\label{sec:model}
\begin{figure}[!htb]
\centering
\includegraphics[width=.66\linewidth]{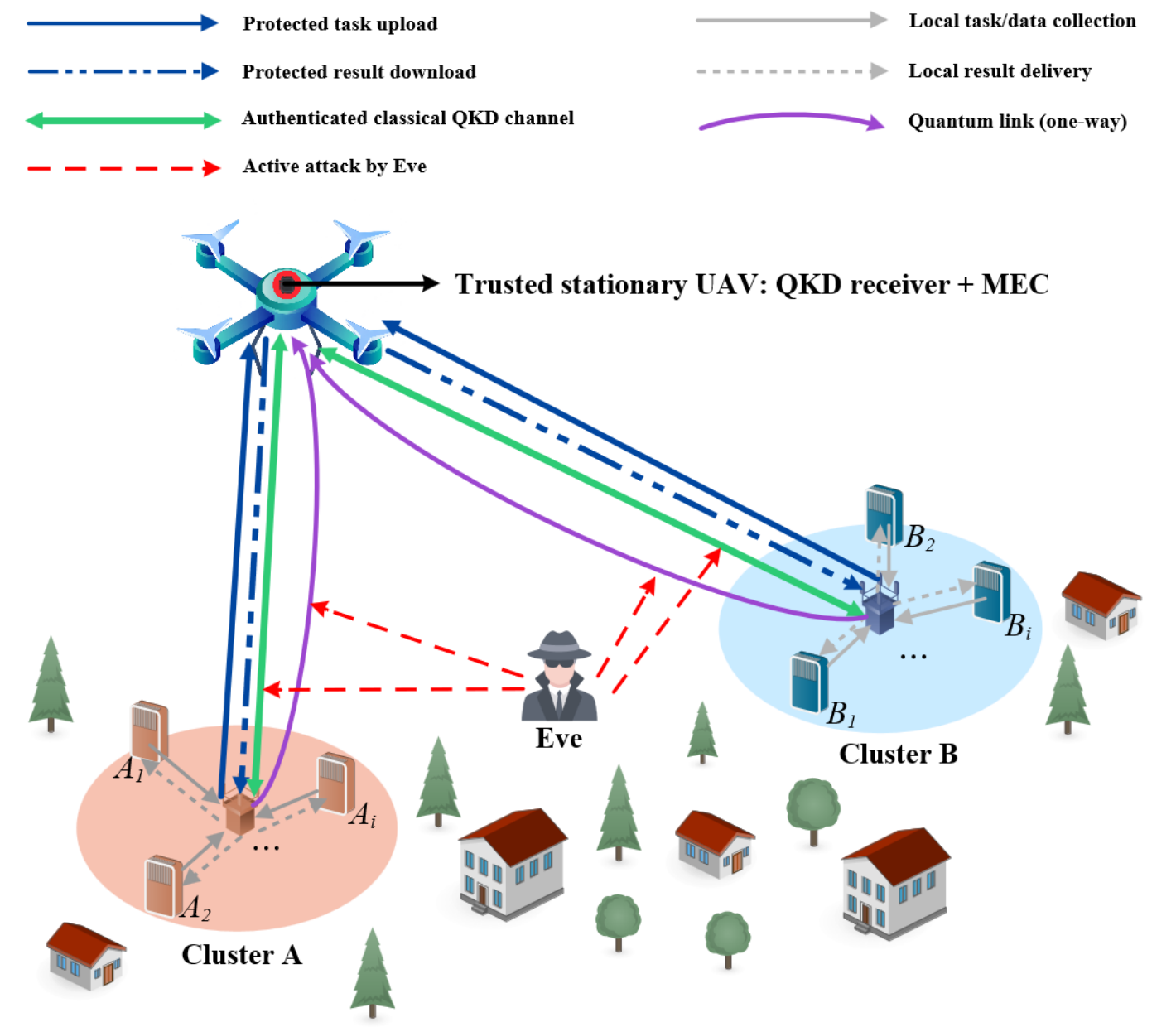}
\caption{Two-cluster system. A trusted stationary UAV hosts the QKD receiver and the MEC server for gateways A and B, serving A first and then B through its single optical receiver. Solid and dash-dot arrows: protected task upload and result download over RF; double-headed arrows: authenticated classical QKD channel; curved arrows: one-way optical quantum uplink ($L_Q=26$~dB); red dashed arrows: active attacks by Eve on the quantum uplink (error injection, extra loss) and on the classical channel (forgery, delay, dropping), Sec.~\ref{sec:threat}; grey arrows: local traffic inside a cluster, outside the timing boundary.}
\label{fig:system}
\end{figure}
\subsection{Topology and service boundary}
\label{sec:topology}
Figure~\ref{fig:system} contains one trusted stationary UAV and two gateways, A and B. Each gateway aggregates its cluster into one task available at $t=0$; local collection and delivery are excluded. Each gateway sends optical quantum signals to the UAV, a separate bidirectional authenticated classical channel carries QKD post-processing, and application data use RF. Service is sequential, A then B, because both stations share the UAV's single optical receiver, without retries, preemption, or concurrency (Sec.~\ref{sec:sens} evaluates an overlapping variant). On acceptance, the UAV decrypts the uploaded input, computes the task, and protects the result; a failed attempt prevents the service but still occupies time.

Each task uses 512 final bits for two AES-256-GCM keys~\cite{nist}. Every attempt burns $a=1440$ pre-shared authentication bits: eighteen 72-bit pads (twelve post-processing and six acquisition messages) and two retired 72-bit selectors, with no recycling across attempts. We use Wegman--Carter authentication~\cite{wegman1981,portmann} with $h_k(x)=\sum_{i=1}^{d}x_i k^i$ over $\mathrm{GF}(2^{72})$, injective length encoding, authenticated indices, fresh pads, message caps, and abort on rejection. The non-constant hash-difference polynomial has at most $d$ roots~\cite[Thm.~2.4]{bibak2022}, giving forgery probability at most $d/2^{72}$ per message; construction and derivation are in the supplementary material. The capped union bound is below $4\times10^{-14}$ per link attempt, giving total security error about $2.03\times10^{-12}$; errors add across attempts under the composition assumptions of~\cite{a11,portmann}.

\subsection{Scope and assumptions}\label{sec:scope}
The UAV is stationary (trust: Sec.~\ref{sec:threat}). The 26-dB effective loss includes coupling and detection efficiency; it combines a measured 17.8-dB drone link at 200~m~\cite{tian} with an assumed 8-dB allowance for unmodeled pointing and turbulence~\cite{scriminich}. RF geometry is separate: this is an uncalibrated composite scenario with night-time background. Acquisition probability 0.99 is an author assumption; all reported reliabilities depend on it. The common processing capacity is an accounting parameter, EC uses a verification proxy, and RF assumes ideal rate adaptation without retransmission. Section~\ref{sec:limits} discusses these limitations.

\subsection{Threat model}
\label{sec:threat}
The service has four security goals. (G1)~Key secrecy: keys come only from a protocol that is $\varepsilon$-secure with $\varepsilon=2\times10^{-12}$~\cite{a11} (Sec.~\ref{sec:qkd}). (G2)~Message authenticity: a forged or substituted post-processing message is accepted with probability below $4\times10^{-14}$ per link attempt (Sec.~\ref{sec:topology}). (G3)~Timely service: both tasks finish with $P_2(D)\geq0.95$ at $D=35$~s. (G4)~Pool sustainability: each pool has drift $\beta\geq0$, and the probability that either link cannot fund one of 100 attempts is at most 0.01, assuming independent, identically distributed acceptance (Sec.~\ref{sec:screen}). G1 and G2 hold up to these errors against any Eve within the trust assumptions below; G3 and G4 are availability goals, which no protocol can guarantee against a channel adversary, and are what we evaluate.

The assets to protect are the application keys (512 bits per task), the authentication-key pool (14,400 bits per link initially, $a=1440$ bits burned per attempt), the task inputs and results, the deadline budget $D$, and the hover energy $E$ of Eq.~\eqref{eq:energy}, which failed attempts also consume. Eve is external and acts on every inter-node link (Table~\ref{tab:threat}). Following Kerckhoffs' principle, she knows the configurations, acceptance thresholds, service order, and deadline, but not the keys, pads, or selectors; she is computationally unbounded against QKD and authentication, and only the application payload relies on computational security. Given G1 and G2, her aim is denial of service (G3, G4).

\begin{table}[!tb]
\caption{Threat model (Sec.~\ref{sec:threat}). ``Assumed'', ``argued'', and ``out of scope'' entries are premises of the model, not results.}
\label{tab:threat}\centering\footnotesize
\setlength{\tabcolsep}{2.5pt}
\begin{tabularx}{\linewidth}{@{}p{.15\linewidth}Yp{.075\linewidth}p{.075\linewidth}Yp{.12\linewidth}@{}}\toprule
Channel / component & Eve's capability & Threat & Goal & Protection / guarantee & Treated in\\\midrule
Quantum uplink & Arbitrary attack on secrecy & T1 & G1 & Finite-key acceptance; $\varepsilon=2\times10^{-12}$~\cite{a11} & Sec.~\ref{sec:qkd}\\
 & Error injection $f$; extra loss $\Delta L$ & T3, T4 & G3, G4 & Acceptance test aborts; protocol $\varepsilon$-security preserved under proof assumptions & Sec.~\ref{sec:eve}, Fig.~\ref{fig:eve}\\
Acquisition beacon & Block acquisition & T5 & G3, G4 & Outside the model (acquisition assumed independent of Eve); if blocked, as T5: 60-s timeout, abort & Argued only\\
Classical QKD channel & Read messages & None & --- & Public by design & ---\\
 & Modify or forge messages & T2 & G2 & One-time Wegman--Carter tags; forgery $<4\times10^{-14}$ per attempt & Sec.~\ref{sec:topology}\\
 & Delay or drop messages & T5 & G3, G4 & Abort; the attempt's $a=1440$ bits are burned & Argued only\\
RF application link & Eavesdrop & --- & Task data & AES-256-GCM~\cite{nist}; computational only & Out of scope\\
 & Jamming & --- & G3 & None & Out of scope\\
UAV, gateways & Compromise & --- & --- & Trusted & Out of scope\\
QKD devices & Side channels & --- & G1 & Devices assumed to match~\cite{a11} & Out of scope\\\bottomrule
\end{tabularx}
\end{table}

Table~\ref{tab:threat} maps each capability to five threats: (T1)~learning the final key; (T2)~forging or substituting a classical message; (T3)~suppressing acceptance so that keys miss the deadline; (T4)~draining the pool through failed attempts; (T5)~blocking or dropping messages so that the attempt aborts. Protocol security stays bounded by $\varepsilon$ under the proof assumptions~\cite{a11}, but attacks may change $p_{\rm pass}$ and hence the conditional bound $\varepsilon/p_{\rm pass}$ for an accepted block (Sec.~\ref{sec:qkd}). We test T3 and T4 on fixed configurations, attacking both stations or only A, with two non-optimal disturbances: error injection $e_f=(1-f)e+f/4$ for both intensities, where $f$ is a matched-basis error strength, not literal intercept-resend on coherent pulses (which would also change yields), and extra loss $\Delta L$, $\eta\mapsto\eta10^{-\Delta L/10}$. They measure the margin of a configuration, not an optimal attack (Sec.~\ref{sec:eve}). Because Eve is unbounded, the thresholds found are upper bounds on robustness: an optimal or adaptive Eve needs at most these disturbances to defeat G3 and G4. Dropping messages (T5) is cheaper still, but it shows on the authenticated channel as a missing or late message. T3 and T4 show only as failed acceptance tests: a 0.2-dB extra loss is small against the 8-dB allowance the model reserves for pointing and turbulence (Sec.~\ref{sec:scope}), and a 0.24-point rise in the error rate is small against the 2.2--6.9\% measured on drone links~\cite{tian,B8_Drone-_and_Vehicle-Based_Quantum_Key_Distribution}. From one attempt's statistics, the service cannot tell such an attack from a poorer channel.

The model rests on the following trust assumptions. The UAV and gateways are trusted, including plaintext access; the QKD devices match the model of~\cite{a11}; authentication keys are provisioned securely before the sortie; local randomness (pads, selectors, basis and sampling choices) is ideal; and AES-256-GCM is computationally secure~\cite{nist}. Acquisition, with success probability $p_{\rm acq}$, is assumed independent of Eve; if she blocked the beacon, the effect would be that of T5: abort after the 60-s timeout, costing time, hover energy, and the attempt's authentication bits, but disclosing no key.

Adaptive or schedule-aware attacks, calibration poisoning, device side channels, platform compromise, and RF jamming are outside the scope of this paper. In particular, the pool-failure bound of G4 does not hold against an adversary who chooses when to disturb the links.
\subsection{Finite-key acceptance with parameters fixed in advance}
\label{sec:qkd}
A configuration fixes, before any sampling, the pulse budget $N$, two positive intensities $\mu_1>\mu_2$ and their probabilities, the basis probabilities, the retained sample sizes $N_Z,N_X$, the acceptance thresholds, the error-correction leakage, and the output length $\ell$. We model phase-randomized weak coherent pulses with active basis choice and one decoy, the protocol of~\cite{a11}. Following~\cite[Eqs.~(73),~(123)]{a11}, with the phase-error deviation
\begin{equation}
\gamma=\sqrt{\frac{s_{Z,1}^{\mathrm l}+s_{X,1}^{\mathrm l}}{s_{Z,1}^{\mathrm l}s_{X,1}^{\mathrm l}}\,
\frac{s_{X,1}^{\mathrm l}+1}{s_{X,1}^{\mathrm l}}\,
\ln\frac{15}{\varepsilon'_{\mathrm{sec}}}},
\end{equation}
the output length is
\begin{equation}
\ell=\Big\lfloor s_{Z,0}^{\mathrm l}+s_{Z,1}^{\mathrm l}
[1-h(\Lambda_X^{\mathrm u}+\gamma)]-\mathrm{leak}_{\mathrm{EC}}
-\log_2\frac{2}{\varepsilon_{\mathrm{cor}}}
-4\log_2\frac{15}{\varepsilon'_{\mathrm{sec}}\,2^{1/4}}\Big\rfloor ,
\label{eq:key}
\end{equation}
where $s_{Z,0}^{\mathrm l}$ and $s_{Z,1}^{\mathrm l}$ are lower thresholds on vacuum and single-photon key-basis events, $s_{X,1}^{\mathrm l}$ on single-photon test events, $\Lambda_X^{\mathrm u}$ is the upper threshold on the single-photon test error rate, and $h$ is binary entropy, set to one above $1/2$. We use $\varepsilon_{\mathrm{cor}}=\varepsilon'_{\mathrm{sec}}=10^{-12}$; by~\cite[Thm.~2]{a11}, the protocol that either outputs $\ell$ bits or aborts is then $\varepsilon$-secure in the sense of~\cite[Def.~1]{a11} with $\varepsilon=\varepsilon_{\mathrm{cor}}+\varepsilon'_{\mathrm{sec}}=2\times10^{-12}$, where Eq.~\eqref{eq:key} sets the internal error terms to the common value $\varepsilon'_{\mathrm{sec}}/15$~\cite[Eq.~(122)]{a11}. Here $\varepsilon'_{\mathrm{sec}}$ bounds the pass probability times the distance from an ideal key~\cite[Eq.~(20)]{a11}, so $\varepsilon$ describes the protocol, not an accepted block~\cite[Sec.~2.3.3]{a11}; for an accepted block only the weaker bound $\varepsilon/p_{\rm pass}$ follows, and Eve influences $p_{\rm pass}$.

For each $N$, the thresholds are derived from the expected detection statistics at the design point with explicit margins: $N_Z$ and $N_X$ are 95\% of the expected sifted counts; $s_{Z,1}^{\mathrm l}$ and $s_{X,1}^{\mathrm l}$ are 90\% of the expected decoy-state lower bounds; $\Lambda_X^{\mathrm u}$ is the expected upper bound plus a \emph{phase margin} of 0.02; the error-correction budget is the expected bit-error fraction plus an \emph{EC margin} of 0.01, and $\mathrm{leak}_{\mathrm{EC}}=1.16\,N_Z\,h(\text{budget})$. The intensities and probabilities are chosen once, at 26~dB and $e=1\%$, by minimizing the $N$ that supplies 512 bits, and are kept for all $N$; re-optimizing them for each $N$ could only raise $\ell(N)$, so the replenishment floor below is conservative (by 1.5\% in the semi-analytic model of Sec.~\ref{sec:semi}), whereas the deadline ceiling may move either way, since the source changes the detection count and hence the hashing workload. Section~\ref{sec:loss} re-chooses them for every loss and misalignment. A trial is accepted if the counts supply $N_Z,N_X$ retained samples, the retained $Z$ error fraction is within the EC budget (error correction is then assumed to succeed), and the decoy bounds satisfy the thresholds; it then releases exactly $\ell$ bits. Thus $\ell\geq512$ alone is not acceptance.

\subsection{Optical counts and QKD timing}
\label{sec:counts}
Multinomial detection counts and hypergeometric retained subsets use the correct/wrong-detector click model specified in the supplementary material; double clicks are assigned at random and afterpulsing is omitted.

Acquisition takes 1~s and succeeds with probability $p_{\rm acq}$; a timeout takes 60~s and ends the attempt. Optical collection takes $N/f_{\rm rep}$. Post-processing exchanges the twelve messages of the supplementary message ledger; an attempt that fails for insufficient samples, error-correction verification, or the acceptance test stops after 2, 6, or 8 messages. For an executed prefix of $J$ messages,
\begin{equation}
T^{\rm post}=\frac{\sum_k W_k}{v_Q}+n_W t_0+
\frac{\sum_{j=1}^{J}m_j}{R_Q}+J t_c,
\label{eq:post}
\end{equation}
where $W_k$ are the declared workloads of the executed operations (bits processed), $n_W$ their number, $t_0=1$~ms a start-up time per operation, $m_j$ the wire length of message $j$ (payload, 256-bit header, and 72-bit tag), $R_Q=10$~Mbit/s the control-channel rate, and $t_c=1$~ms the one-way control latency. Both endpoints are serialized. Every message is hashed at both endpoints for authentication, including the 128-bit length prefix that makes the encoding injective (Sec.~\ref{sec:model}). Because the first message announces the indices of all detections, it alone carries 64\% of the hashing workload (the supplementary message ledger), which makes authentication the largest variable cost. The acceptance test is charged its $N_X$ comparisons, since its bounds use aggregate counts.

\subsection{RF, computation, completion time, and hover energy}
Inputs are 40 and 80~Mbit, results half as large; computation takes 2 and 4~s (Table~\ref{tab:inputs}). AES-GCM timings interpolate the IoT profiles of~\cite{prisec}; each direction adds a 96-bit IV and a 128-bit tag~\cite{nist}. RF transfer time is payload divided by $B_{\rm w}\log_2(1+\bar\rho H)$, with mean SNR $\bar\rho=877.2$ (29.4~dB), independent $H\sim\operatorname{Gamma}(2,1/2)$ per transfer~\cite{E5_The_m-Distribution_A_General_Formula_of_Intensity_Distribution_of_Rapid_Fading}. Geometry and full equations are in the supplementary material.
Let $\chi_i\in\{0,1\}$ indicate that station $i$'s attempt is accepted. A successful attempt takes
\begin{equation}
S_i=T_i^{\rm acq}+N/f_{\rm rep}+T_i^{\rm post}
+T_i^{\rm cry}+T_i^\uparrow+T_i^{\rm MEC}+T_i^\downarrow .
\end{equation}
The QKD part $Q_i=T_i^{\rm acq}+N/f_{\rm rep}+T_i^{\rm post}$ is the time before keys exist. The actual duration $U_i$ equals $S_i$ on success and the executed prefix on failure. With deadlines measured from common arrival,
\begin{equation}
C_A=U_A\ \text{if }\chi_A=1,\qquad
C_B=U_A+U_B\ \text{if }\chi_B=1,
\label{eq:completion}
\end{equation}
and $+\infty$ otherwise. Both tasks finish on time when $\max(C_A,C_B)\leq D$; a failure of A delays B even if B succeeds.

The UAV hovers from $t=0$ until the round ends, including failed attempts, so a service round costs the hover energy
\begin{equation}
E=P_h\,(U_A+U_B),
\label{eq:energy}
\end{equation}
with hover power $P_h=1.6$--$1.7$~kW. For a stationary UAV, the rotary-wing model of~\cite{E6_Energy_Minimization_for_Wireless_Communication_with_Rotary-Wing_UAV} reduces to hovering power; we take $P_h$ from a measured drone carrying a QKD payload~\cite{B8_Drone-_and_Vehicle-Based_Quantum_Key_Distribution} and neglect radio, QKD, and computing power (tens of watts~\cite{B8_Drone-_and_Vehicle-Based_Quantum_Key_Distribution}). Energy is a metric, not a constraint.
\section{Metrics and Experimental Setup}
\label{sec:setup}
\begin{table}[!htb]
\caption{Simulation inputs. R: setting from the cited reference; D: derived; A: author-defined assumption or design choice. The inputs form a composite benchmark, not one calibrated platform.}
\label{tab:inputs}\centering\small
\begin{tabularx}{\linewidth}{@{}p{.31\linewidth}Y@{}}\toprule
Optical loss; pulse rate & 26~dB per station (17.8~dB measured in~\cite{tian}), varied 18--30~dB; $f_{\rm rep}=50$~MHz, also 100~MHz (A)\\
Source and basis & $(\mu_1,\mu_2)=(0.327,0.136)$, $p_{\mu_1}=0.357$, $p_Z=0.544$ at both ends (D, design rule)\\
Detectors & Dark 100~s$^{-1}$ per detector; background 100~s$^{-1}$ (night); gate 2~ns; misalignment $e=1\%$, varied 1--3\% (A)\\
Finite-key security & $\varepsilon_{\rm cor}=\varepsilon'_{\rm sec}=10^{-12}$; $f_{\rm EC}=1.16$; formula~\cite{a11} (R/A)\\
Design margins & Retained fraction 0.95; threshold fraction 0.9; phase margin 0.02; EC margin 0.01 (A)\\
Keys & 512 bits per task; 1,440 authentication bits per attempt (A)\\
Acquisition & Success probability 0.99; 1~s on success, 60-s timeout (A)\\
QKD processing & $v_Q=2$--30~Mworkbit/s, reference 10; $R_Q=10$~Mbit/s; $t_0=t_c=1$~ms (A)\\
Tasks & $I_A=40$, $I_B=80$~Mbit; $O_i=I_i/2$; 1,500 cycles/bit~\cite{li2019} (R)\\
MEC & 30~Gcycles/s~\cite{sun} (R), giving 2 and 4~s (D)\\
RF & 40~MHz; noise $10^{-9}$~W; $g_0=10^{-3}$~\cite{li2019}; 0.2~W~\cite{li2020} (R); UAV at (8,8,10)~m, stations at (0,0,0) and (16,0,0)~m (A); mean SNR 29.4~dB (D); Nakagami $m=2$ (A)\\
Application protection & AES-GCM timing~\cite{prisec}; 224-bit overhead~\cite{nist} (R)\\
Hover power & 1.6--1.7~kW~\cite{B8_Drone-_and_Vehicle-Based_Quantum_Key_Distribution} (R)\\
Deadlines; pool & $D\in\{30,35,40\}$~s; initial pool 14,400 bits per link; 100-attempt horizon (A)\\\bottomrule
\end{tabularx}
\end{table}
\subsection{Configuration families and metrics}
Table~\ref{tab:inputs} lists the inputs. For local availability, the demands 512, 768, and 1,024 bits use fixed retained sizes $(N_Z,N_X)=(40710,28612)$, $(42594,29936)$, and $(44426,31224)$ and fixed thresholds, while only the emitted budget $N$ varies (80 budgets in 250--330 million, refined near each transition; 2,000 trials each). For joint service, 31 budgets span 281.8--563.7 million pulses; for each $N$ the retained sizes, thresholds, leakage, and output are redesigned \emph{before} sampling, and both stations use the same configuration. Representatives are $(N/10^6,\ell)=(310.0,1074)$, $(357.0,2077)$, $(422.7,3592)$, and $(563.7,7164)$.

Local availability is $\widehat P_k(N)=n^{-1}\sum_r\ind\{\chi_r=1,\ell\geq k\}$. With $C_r=\max(C_{A,r},C_{B,r})$ for trial pair $r$, the joint reliability and the empirical 95th completion percentile are
\begin{equation}
\widehat P_2(D;v)=\frac1n\sum_r\ind\{C_r(v)\leq D\},\qquad
\widehat D_{95}(v)=C_{(\lceil0.95n\rceil)}(v),
\label{eq:quantile}
\end{equation}
with failed pairs at infinity, so $\widehat D_{95}=\infty$ whenever $\widehat P_2<0.95$ for every $D$. For QKD duration $Q$ and executed workload $W$, a change of capacity reuses the same outcomes and changes only variable processing time:
\begin{equation}
Q_{i,r}(v)=Q_{i,r}(v_0)+W_{i,r}(1/v-1/v_0),\quad v_0=10^7 .
\label{eq:scaling}
\end{equation}
The gross accepted-key throughput, with failures contributing time and no output, is
\begin{equation}
\widehat R_{\rm key}(v)=\frac{\sum_{i}\sum_r \chi_{i,r}\ell}{\sum_{i}\sum_r Q_{i,r}(v)} ;
\label{eq:throughput}
\end{equation}
it does not subtract authentication or application key consumption, and its form $K/(c_0+c_1/v)$ explains diminishing returns.

\subsection{Monte Carlo procedure and semi-analytic model}
\label{sec:semi}
Count-level Monte Carlo generates optical detection statistics, acquisition outcomes, and RF realizations; cryptographic processing is represented by its acceptance and cost models. The availability sweep generates 480,000 local blocks and the joint family 186,000; the first 1,000 of 3,000 pairs per configuration are reserved for design. Acquisition and RF streams are shared across configurations (common random numbers). The statistics of Secs.~\ref{sec:budget}--\ref{sec:lossres} were recomputed independently from the stored samples, except the rate-model, hash-workload, source re-optimization, and hashing-benchmark analyses; of Sec.~\ref{sec:eve}, only the key probabilities with both links attacked (Fig.~\ref{fig:eve}) were.

Because completion time given acceptance is nearly deterministic, a semi-analytic model reproduces the simulation. It assumes (A1) independent pulses with per-pulse click probability $q=\sum_k p_{\mu_k}[1-(1-q_c)(1-q_w)]|_{\mu_k}$; (A2) sifted totals $\mathrm{Bin}(N,p_Z^2q)$ and $\mathrm{Bin}(N,(1-p_Z)^2q)$, treated as independent (their multinomial correlation $\approx-p_Z(1-p_Z)q$ is below $10^{-4}$ in magnitude at 26~dB); (A3) conditional on sufficient counts, retained $Z$ errors $\mathrm{Bin}(N_Z,\epsilon_Z)$ with the per-click error probability $\epsilon_Z$; (A4) the decoy and phase-error thresholds hold whenever the counts suffice and error correction passes; (A5) acquisition independent of everything else; and (A6) latency equal to its value at the expected detection count plus the RF fading term. The two stations are independent, and RF fading is independent of acceptance; these are model assumptions. Then the local acceptance probability is
\begin{equation}
\begin{split}
p_{\rm loc}=p_{\rm acq}&\Pr[\mathrm{Bin}(N,p_Z^2q)\geq N_Z]\,\Pr[\mathrm{Bin}(N,(1-p_Z)^2q)\geq N_X]\\
&\times\Pr[\mathrm{Bin}(N_Z,\epsilon_Z)\leq\lfloor \delta_{\rm EC}N_Z\rfloor],
\end{split}
\label{eq:semi}
\end{equation}
with EC budget $\delta_{\rm EC}$, and $\widehat P_2(D)\approx p_{\rm loc}^2\Pr[T_{\rm det}+T_{\rm fade}\leq D]$, where $T_{\rm det}$ is the deterministic part and $T_{\rm fade}$ the sum of the four RF transfer times. Assumption (A4) is supported by the simulation: in the joint family, acceptance-test failures occur in 0.04\% of acquired blocks (at most 0.22\% for one configuration), on the undisturbed channel only.

For one family (fixed source, channel, and design rule) and integer $N$ in its stored range, the model bounds the window off the grid (derivation in the supplementary material). As $\beta\geq0$ iff $\ell\geq\lceil512+a/p_{\rm loc}\rceil$ and Chernoff bounds on~\eqref{eq:semi} give an explicit $\underline p(N_0)\leq p_{\rm loc}(N)$ for $N\geq N_0$, $\ell\geq Q=\lceil512+a/\underline p\rceil$ (1,967 bits) suffices. Let $T(N)=T_{\rm det}(N)+q(N)$, with $q$ the $0.95/p_{\rm loc}^2$ quantile of $T_{\rm fade}$, so $T(N)\leq D$ is the model's $\widehat P_2(D)\geq0.95$.
\begin{proposition}\label{prop:window}
For the fixed proportional-retention family and admissible stored domain, under (A1)--(A6) and the independence assumptions above: (i) if $N^*$ is the smallest $N$ at which an expected-count lower bound $rN-b\sqrt N-c\leq\ell(N)$ reaches $Q$, then $\beta\geq0$ for every admissible $N\geq N^*$; (ii) between jumps of $\lceil\log_2N\rceil$, $T_{\rm det}$ lies between two affine functions of $N$ of equal slope, which give integers $N^-\leq N^+$ with $T\leq D$ on $[N^*,N^-]$ and $T>D$ above $N^+$. Hence $N^*\leq N^-$ implies a non-empty window.
\end{proposition}
The condition is sufficient only. At 26~dB and $e=1\%$, $N^*=352.04$ million (grid floor 357.0); at 10~Mworkbit/s the ceiling is about 409.9 million ($\kappa=1$) and 496.8 million ($\kappa=0$) at 35~s and 386.4 million ($\kappa=0$) at 30~s ($N^+-N^-\leq1{,}069$ pulses), and snaps to the windows of Table~\ref{tab:kappa} (its empty $\kappa=1$, 30-s window remains a grid result). In 701 of the 702 loss-sweep combinations the snapped bound equals the simulated grid ceiling (the exception, at 18~dB, is one grid step off), and all loss limits match; as $q$ uses the stored fading samples, this checks the approximations, not the fading model.

\subsection{Feasibility screen and held-out confirmation}
\label{sec:screen}
An attempt requires at least $a=1440$ authentication bits. The full-burn pool evolves as

\begin{equation}
A_{t+1}=A_t-a+\chi_t(\ell-512),\qquad A_0=14400,
\label{eq:inventory}
\end{equation}
rather than a recycling protocol~\cite{portmann}. Its drift is $\beta=p_{\rm loc}(\ell-512)-a$. Nonnegative drift is a policy requirement for a persistent pool, not a security condition; refilling at landing can remove it. Charging only executed pads would not move the current grid floor: accepted attempts still consume all 1,440 bits. At 10~Mworkbit/s, design pairs select the smallest grid budget satisfying both $\widehat P_2(D)\geq0.95$ and $\beta\geq0$. A new seed then supplies 5,000 held-out pairs for that choice, the deadline ceiling, and two controls. Confirmation requires

\begin{equation}
\underline P_2(D)\geq0.95,\qquad \min_i\{\underline p_i(\ell-512)-a\}\geq0,
\label{eq:jointscreen}
\end{equation}
and an upper bound of at most 0.01 on the probability that either link cannot fund one of 100 attempts, computed exactly from Eq.~\eqref{eq:inventory} with $\underline p_i$ and a union bound over links, assuming independent, identically distributed acceptance. Here $\underline P_2$ and $\underline p_i$ are one-sided exact binomial lower bounds with Bonferroni tail $0.05/80$: four configurations times (three capacities $\times$ three deadlines $\times$ two workload profiles $+$ two local probabilities). The profiles are nominal and a stress in which the authentication-hashing and acceptance-test workloads double. Neither statistical coverage nor the pool criterion is cryptographic composition (Sec.~\ref{sec:model}).

\subsection{Sensitivity, baselines, disturbance, and channel sweep}
\label{sec:sensmethod}
We vary authentication debit, hashing cost, UAV-side encryption, pulse rate, acquisition, retries, and scheduling on stored samples. Baselines choose the smallest key-sufficient budget (B1), maximum gross throughput (B2), or smallest timely budget without the balance constraint (B3). Disturbance uses 4,000 pairs per level; the margin sweep uses 3,000. Full grids and procedures are in the supplementary material.

\subsection{Optical loss and misalignment}\label{sec:loss}
For each loss and misalignment, the source is re-chosen and 41 budgets are redesigned and sampled with 2,000 pairs each: 22--30~dB at $e=1,2,3\%$, and 18--21~dB at $e=1\%$. A separate-seed exploratory grid adds $e=2,3\%$ at 18--21~dB. Ceilings below or beyond the stored grid are flagged. The window runs from the smallest grid budget with nonnegative balance at both stations to the largest meeting $\widehat P_2(D)\geq0.95$.
\FloatBarrier
\section{Numerical Results}
\label{sec:results}
\begin{figure}[!tb]
\centering\includegraphics[width=0.9\linewidth]{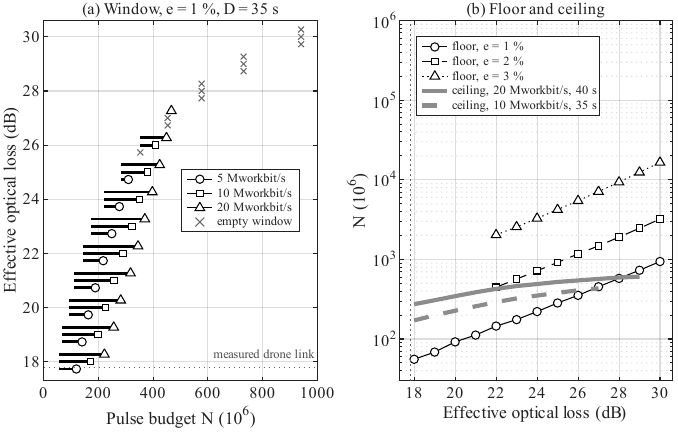}
\caption{High-cost stress reference ($\kappa=1$); loss limits with benchmarked hashing are in Table~\ref{tab:kappa}. (a) Feasibility window $[N_{\min},N_{\max}]$ versus effective loss at $e=1\%$ and $D=35$~s for three capacities (bars end at the ceiling; crosses mark an empty window at the floor); the dotted line marks the 17.8~dB measured in~\cite{tian}. (b) Replenishment floor for $e=1$, 2, and 3\% (the last two drawn only for 22--30~dB) against two deadline ceilings computed at $e=1\%$ (20~Mworkbit/s, 40~s; 10~Mworkbit/s, 35~s); windows at $e=2$, 3\% use the ceiling at the same $e$ (not drawn; Table~\ref{tab:kappa}). Source and configuration are redesigned for every loss and misalignment; 2,000 pairs per configuration.}
\label{fig:loss}
\end{figure}

Unless stated otherwise, results use 26~dB, $e=1\%$, 10~Mworkbit/s, and $D=35$~s, on the 2,000 evaluation pairs per configuration (held-out estimates differ slightly). Headline results use proxy-benchmarked hashing ($\kappa\approx0$, computed at $\kappa=0$; Sec.~\ref{sec:sens}; Table~\ref{tab:kappa}). Figures~\ref{fig:latency}--\ref{fig:loss} and Tables~\ref{tab:budget}--\ref{tab:decision} show the high-cost stress reference ($\kappa=1$), which charges hashing in full; the floor is the same for both.

\subsection{Where the time and the energy go}
\label{sec:budget}
\begin{table}[!tb]
\caption{Mean joint completion time of both tasks by component (reference settings; pairs in which both stations succeed); the components add up exactly to each pair's completion time. Hover energy ($P_h=1.6$~kW) includes failed attempts.}
\label{tab:budget}\centering\small
\setlength{\tabcolsep}{5pt}
\begin{tabular}{@{}lrrr@{}}\toprule
Pulse budget $N$ (million) & 310.0 & 357.0 & 563.7\\
Output $\ell$ (bits) & 1074 & 2077 & 7164\\\midrule
Acquisition (s) & 2.00 & 2.00 & 2.00\\
Optical collection (s) & 12.40 & 14.28 & 22.55\\
Authentication hashing (s) & 2.98 & 3.43 & 5.59\\
Other QKD processing (s) & 0.23 & 0.25 & 0.36\\
Control-channel transfer (s) & 1.51 & 1.74 & 2.82\\
AES-GCM, both ends (s) & 3.81 & 3.81 & 3.81\\
MEC computation (s) & 6.00 & 6.00 & 6.00\\
RF data transfer (s) & 0.49 & 0.49 & 0.49\\\midrule
Mean total (s) & 29.42 & 31.99 & 43.60\\
$\widehat D_{95}$ (s) & 29.52 & 32.10 & 43.71\\
$\widehat R_{\rm key}$ (bit/s) & 102.6 & 176.7 & 408.2\\
Net key balance $\beta$ (bits/attempt) & $-887$ & 101 & 5112\\
Hover energy per round (kJ) & 49.2 & 53.3 & 71.6\\
\quad share before keys exist & 67\% & 70\% & 77\%\\
Energy per on-time task (kJ) & 25.2 & 27.2 & 72.7\\\bottomrule
\end{tabular}
\end{table}
Pre-existing application keys give $\widehat D_{95}=10.37$~s; a finite stock serves only finitely many rounds. Fresh keys raise this to 32.10~s for the selected stress-reference configuration, of which 70\% of the 53.3~kJ hover energy precedes key release (Table~\ref{tab:budget}). ML-KEM-768 would take about 2~ms of control transmission plus computation~\cite{D9_Module-Lattice-Based_Key-Encapsulation_Mechanism_Standard}, leaving completion near 10.37~s, under computational security and separate authentication. Gross scalar rates predict 16.0 or 10.8~s instead of 32.1~s. Charging the authentication debit gives 29.2~s, but only 13.5~s for the largest block versus 43.7~s: a rate cannot represent the wait for whole-block release. Optical collection dominates; hashing exceeds other QKD processing only under the assumed stress workloads. Detailed comparisons are in the supplementary material.

The semi-analytic model of Sec.~\ref{sec:semi} reproduces these numbers: over the 31 configurations and three capacities, $\widehat D_{95}$ differs by at most 0.027~s and the success fraction given acquisition by at most 0.005. The simulated $\widehat P_2$ lies about one point below $p_{\rm acq}^2=0.980$ because the shared acquisition stream acquired both stations in 97.1\% of evaluation pairs (98.2\% held out); as completion given acceptance is nearly deterministic, $\widehat P_2(D)$ is close to a step between 0.98 and 0.

\subsection{Pulse budget and local key availability}
\begin{figure}[!tb]
\centering\includegraphics[width=.55\linewidth]{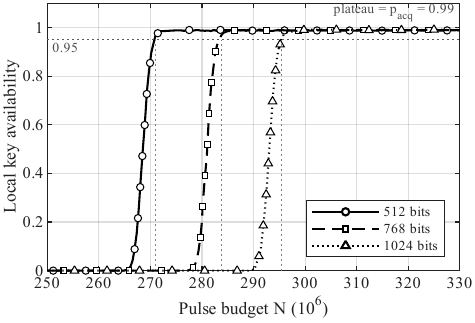}
\caption{Local key availability at 26~dB. Within each demand, retained samples and thresholds are fixed while $N$ varies; acquisition failures are included. Monte Carlo estimates; the dotted vertical lines mark the first budget at which each curve reaches 0.95.}
\label{fig:availability}
\end{figure}
Higher key demand shifts the transition to larger budgets (Fig.~\ref{fig:availability}). The first empirical 0.95 crossings on the grid are 270.94, 283.73, and 295.38 million pulses for 512, 768, and 1,024 bits; the semi-analytic model~\eqref{eq:semi} places them at 270.77, 283.23, and 295.33 million, within 0.14 million of the interpolated simulation crossings. On 6,000 fresh trials per crossing, the estimates are 0.9548, 0.9692, and 0.9490 (the last below 0.95; 95\% interval $[0.943,0.954]$).

\subsection{Processing capacity, key throughput, and latency}
\begin{figure}[!tb]
\centering
\includegraphics[width=.84\linewidth]{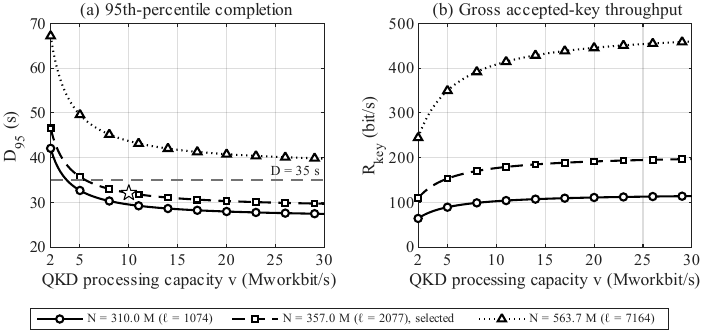}
\caption{Effect of QKD processing capacity for three configurations at 26~dB. (a) Empirical 95th-percentile joint completion time; the dashed line marks the 35-s deadline and the star the selected configuration at the reference capacity of 10~Mworkbit/s ($\widehat D_{95}=32.10$~s). (b) Gross accepted-key throughput~\eqref{eq:throughput}, including failed attempts.}
\label{fig:latency}
\end{figure}
Capacity shortens processing with diminishing returns but cannot remove optical collection (Fig.~\ref{fig:latency}a, for 310.0, 357.0, and 563.7 million pulses): for the selected configuration, $\widehat D_{95}$ is 35.71, 32.10, and 30.30~s at 5, 10, and 20~Mworkbit/s. Larger blocks are slower at every capacity, yet they yield more accepted bits per second of QKD service (Fig.~\ref{fig:latency}b). Maximizing gross key throughput therefore selects the configuration that serves a fixed-demand, deadline-constrained task worst. Positive throughput also differs from sustainable key replenishment: with $\ell=1074$, a successful attempt still loses $1440-(1074-512)=878$ pool bits, so a 14,400-bit pool funds at most 15 attempts.

\subsection{The feasibility window and its confirmation}
\label{sec:decision}
\begin{figure}[!tb]
\centering\includegraphics[width=0.78\linewidth]{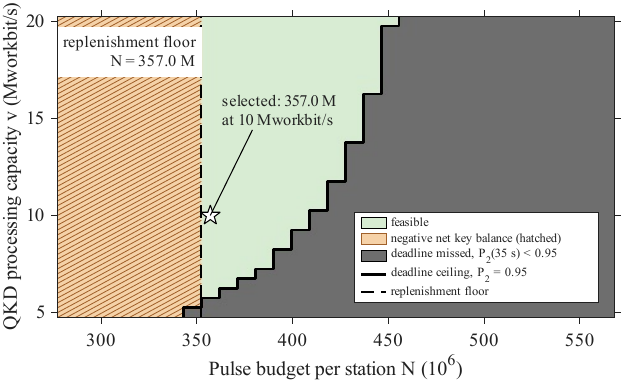}
\caption{Deadline--replenishment feasibility at $D=35$~s on 31 configurations and 31 capacities, in the high-cost stress reference ($\kappa=1$); with benchmarked hashing ($\kappa\approx0$) the floor is unchanged and the ceiling rises (Table~\ref{tab:kappa}). Hatched: negative net key balance; dark: $\widehat P_2(35)<0.95$; light: feasible; dashed: replenishment floor; stepped: deadline ceiling; star: selected configuration. Region boundaries are drawn at grid-cell edges; labelled values (the floor label is the first feasible grid value, 357.0 million) and the star refer to grid configurations.}
\label{fig:resource}
\end{figure}
Figure~\ref{fig:resource} shows the central result. Two requirements bound the pulse budget from opposite sides. The \emph{replenishment floor} is set by the key-balance requirement: every tested grid configuration below 357.0 million pulses outputs $\ell\leq1870<1440+512$ bits and cannot replenish its pool even if every attempt succeeds. We report this grid value; the threshold lies in $(347.6,357.0]$ million, and the model's sufficient bound is 352.04 million (Proposition~\ref{prop:window}). The \emph{deadline ceiling} moves with capacity: with benchmarked hashing it is 488.5, 488.5, and 497.9 million pulses at 5, 10, and 20~Mworkbit/s. In the high-cost stress reference it is 338.2, 404.0, and 450.9 million (fresh checks, 6,000 pairs each: 0.970--0.980), so at 5~Mworkbit/s it lies below the floor and the window is empty.

The screen therefore selects the floor, $N=357.0$ million and $\ell=2077$, at 35 and 40~s, and at 30~s only with benchmarked hashing (Table~\ref{tab:kappa}). Within the window, hover energy per on-time task grows with $N$, so the floor is also the most energy-efficient feasible choice; beyond the ceiling, B misses the deadline and the energy per on-time task more than doubles (Table~\ref{tab:budget}).

\begin{table}[!tb]
\caption{Held-out confirmation at 10~Mworkbit/s (new seed, 5,000 pairs per configuration). $\underline P_2$: simultaneous lower bound (Bonferroni, 80 bounds); stress: hashing and acceptance-test workloads doubled; 0: no pair finishes within 35~s. Balance: smaller station lower bound on $\beta$. Pass at 35~s: $\underline P_2\geq0.95$, nonnegative balance, and the pool criterion.}
\label{tab:decision}\centering\small
\setlength{\tabcolsep}{3.5pt}
\begin{tabular}{@{}rrcccccrc@{}}\toprule
&&\multicolumn{3}{c}{Nominal}&\multicolumn{2}{c}{Stress}&&\\\cmidrule(lr){3-5}\cmidrule(lr){6-7}
$N$ ($10^6$)&$\ell$&$\widehat P_2(35)$&$\underline P_2(35)$&$\widehat D_{95}$ (s)&$\underline P_2(35)$&$\widehat D_{95}$ (s)&Balance&Pass\\\midrule
310.0&1074&0.980&0.973&29.50&0.973&32.48&$-887.7$&no (balance)\\
357.0&2077&0.981&0.973&32.08&0&35.52&98.2&yes\\
404.0&3147&0.981&0.974&34.66&0&38.54&1149.9&yes\\
563.7&7164&0&0&43.69&0&49.29&5101.5&no (deadline)\\\bottomrule
\end{tabular}
\end{table}
Table~\ref{tab:decision} confirms the choice on held-out pairs. The selected configuration has $\widehat P_2(35)=0.981$ with simultaneous lower bound 0.973 (given $p_{\rm acq}=0.99$), balance lower bounds of 103.6 and 98.2 bits per attempt, and a pool-failure bound of $7\times10^{-11}$. Under the stress profile it misses 35~s ($\widehat D_{95}=35.52$~s) but passes 40~s; the ceiling configuration (404.0 million) passes at 35~s with 2.6~s less slack.

Simpler rules fail one requirement each (evaluation pairs, $D=35$~s, exact 95\% intervals). The key-length and deadline-only rules (B1/B3, 281.8 million) give $\widehat P_2(35)=0.968$ $[0.959,0.975]$ but a balance of $-1440$ bits, so the pool funds only 10 attempts under full burn; the highest-throughput rule (B2, 563.7 million) gives 0.000 $[0,0.002]$. The joint rule gives 0.970 $[0.962,0.977]$ with a balance of 100.7 bits.

\subsection{What moves the window}
\label{sec:sens}

For a fixed source, channel, acceptance model, and application allocation, varying the authentication debit shifts the replenishment floor, monotonically in $a$ (full debit sweep in the supplementary material); hashing cost does not, and decides only which deadlines the floor meets. The reference $a=1440$ is the allocation of Sec.~\ref{sec:model}: eighteen tag pads and two retired selectors of 72 bits each, with no recycling. The values 1296 and 1008 would follow from reusing hash selectors across attempts and from not charging the acquisition-phase tags, whose security conditions~\cite{portmann} we do not establish. On the held-out pairs, the selected configuration keeps a nonnegative balance bound up to $a=1440$ but not at 2,160 ($-621.8$).

With benchmarked hashing the ceiling barely changes with processing capacity; processing controls it only in the stress reference. Hash-cost factors $\kappa=0,0.5,1,2$ give held-out $\widehat D_{95}=28.65,30.36,32.08,35.51$~s. Poly1305/GHASH proxies process both endpoints' message workload in under 0.3~ms ($\kappa<2\times10^{-4}$); this motivates $\kappa\approx0$ but calibrates neither the specified field authenticator nor UAV execution. Lowering $\kappa$ preserves the held-out reliability lower bound 0.973.

Scheduling overlap and a doubled pulse rate raise the stress-reference ceiling without changing the replenishment policy. Acquisition imposes a separate cap: $P_2\leq p_{\rm acq}^2$, or $p_{\rm acq}^K$ for $K$ independent stations; at $p_{\rm acq}=0.99$, at most five can meet 0.95. Detailed sensitivities and retry results are in the supplementary material.

\subsection{Optical loss and misalignment}
\label{sec:lossres}
\begin{table}[!tb]
\caption{Benchmarked hashing ($\kappa\approx0$, computed at $\kappa=0$) against the high-cost stress reference ($\kappa=1$) at $D=35$~s. Windows in million pulses at 10~Mworkbit/s, design pairs. Loss limit: largest tested loss (18--30~dB, 1-dB steps; at $e=2$, 3\%, 18--21~dB on the exploratory grid) with a non-empty window at 5/10/20~Mworkbit/s; $\emptyset$: none. Held-out rows: new seed, 5,000 pairs per loss, 10~Mworkbit/s, floor configuration; $\underline P_2$: simultaneous lower bound (Bonferroni, 12 claims); balance: smaller station's lower bound on $\beta$ per attempt.}
\label{tab:kappa}\centering\small
\setlength{\tabcolsep}{4pt}
\begin{tabular}{@{}llcc@{}}\toprule
Quantity & Setting & $\kappa\approx0$ & $\kappa=1$\\\midrule
Window & 26~dB, $e=1\%$ & $[357.0,488.5]$ & $[357.0,404.0]$\\
Window at $D=30$~s & 26~dB, $e=1\%$ & $[357.0,385.2]$ & $\emptyset$\\
Held-out $\widehat D_{95}$ (s) & 26~dB, $e=1\%$ & 28.65 & 32.08\\
Loss limit (dB) & $e=1\%$ & 27/27/27 & 25/26/27\\
 & $e=2\%$ & 22/22/22 & $\emptyset$/20/21\\
 & $e=3\%$ & $\emptyset$/$\emptyset$/$\emptyset$ & $\emptyset$/$\emptyset$/$\emptyset$\\\midrule
\multicolumn{2}{@{}l}{Held out, $e=2\%$, 18/19/20~dB: $N_{\rm floor}$ ($10^6$)} & \multicolumn{2}{c}{175.5/223.5/280.5}\\
\quad $\underline P_2(35)$ && 0.969/0.969/0.969 & 0.969/0.969/0.969\\
\quad $\widehat D_{95}$ (s) && 23.34/25.28/27.60 & 30.27/32.24/34.68\\
\quad Min.\ balance bound (bits) &&\multicolumn{2}{c}{130.2/206.7/105.7}\\\bottomrule
\end{tabular}
\end{table}
Table~\ref{tab:kappa} gives the largest tested loss with a non-empty 35-s window. With benchmarked hashing, it is 27~dB at $e=1\%$ and 22~dB at $e=2\%$; no tested setting has a window at $e=3\%$. Twenty-nine $\kappa=0$ ceilings lie beyond the stored range and are lower bounds only. At 26~dB, $e=2\%$, the floor reaches 1,171 million pulses, beyond every ceiling. Source redesign at each $(L_Q,e)$ prevents attributing cross-setting ceiling changes to misalignment alone. Full exploratory-grid results are in the supplementary material.

The held-out check of these floors (Table~\ref{tab:kappa}) passes at 18--20~dB for both $\kappa$, including the pool criterion. All local QKD failures are acquisition failures; one jointly accepted pair misses 35~s at 20~dB under $\kappa=1$, whose window is a single configuration with 0.32~s of slack. The check confirms the floor arithmetic, the timing, and the assumed $p_{\rm acq}^2$ (the three losses share the acquisition stream), not the robustness of finite-key acceptance. A second-seed replication gives $\underline P_2(35)=0.977$ and $\widehat D_{95}$ within 0.01~s of these values. Across the 279 earlier combinations ($\kappa=1$; 22--30~dB; 18--21~dB at $e=1\%$), the semi-analytic screen, with its own fading draws, and the simulation agree in all floors, selections, and the 96 ceilings on the grid; the other 183 lie below it.

\subsection{Availability under disturbance}
\label{sec:eve}
\begin{figure}[!tb]
\centering\includegraphics[width=.72\linewidth]{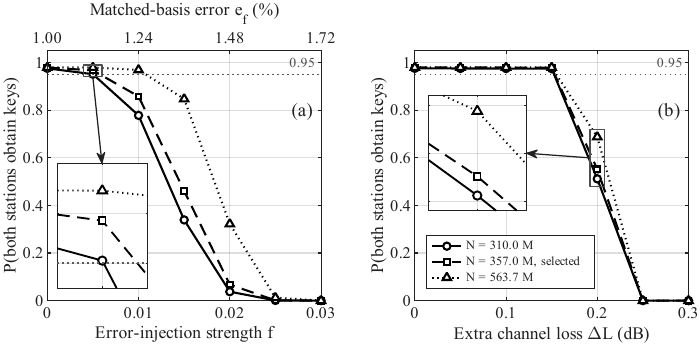}
\caption{Probability that both stations obtain keys under (a) error injection of strength $f$ (top axis: matched-basis error $e_f=(1-f)\,0.01+f/4$) and (b) extra channel loss, both links attacked. Configurations are fixed; 4,000 pairs per point; acquisition failures included. The dotted line marks 0.95; insets magnify the boxed regions (values in the text). Monte Carlo only; (A4) is not validated under disturbance.}
\label{fig:eve}
\end{figure}
Fixed configurations that meet the deadline tolerate only small disturbances (Fig.~\ref{fig:eve}); the largest block (563.7 million) degrades later under error injection but misses the deadline even undisturbed (Table~\ref{tab:decision}). For the selected configuration, $f=0.5\%$ leaves $\widehat P_2(35)=0.967$ (the key probability; the deadline does not bind), but $f=1\%$, which raises the matched-basis error only from 1.00\% to 1.24\%, reduces it to 0.858 (95\% interval $[0.847,0.868]$); $f=1.5\%$ and 2\% give 0.461 and 0.067. Extra loss acts as a threshold: up to 0.15~dB the probability stays at 0.978, at 0.2~dB it is 0.552, and at 0.25~dB it is 0.0003. The two cliffs fail at different tests: at 0.2~dB, 1,965 of the 8,000 attempts abort for insufficient samples, mostly in the $X$ basis; at $f=1\%$, all 511 QKD aborts occur at the acceptance test, 319 (62.4\%) of them violating the phase-error bound and the rest only single-photon bounds, which depend on the error counts through the vacuum bound. Attacking only A gives 0.917 at $f=1\%$ and 0.735 at 0.2~dB, so one disturbed link suffices to break the 0.95 target. Below these thresholds $\widehat D_{95}$ barely changes; beyond them it is infinite: the disturbance removes the service rather than slowing it, while $\varepsilon$ is unchanged (Sec.~\ref{sec:threat}). The critical error, about 1.2--1.4\% (far below the 11\% asymptotic one-way BB84 threshold~\cite{alleaume2014}), leaves blocks sized for latency a margin of only 0.2--0.4 percentage points.

The attack also drains the pool: the selected configuration's balance on the disturbance samples falls from 107.8 to 7.6 bits per attempt at $f=1\%$ and to $-383.6$ at $f=1.5\%$; once acceptance vanishes, a 14,400-bit pool funds only 10 attempts under full-burn accounting (Sec.~\ref{sec:screen}). This is the exhaustion route to denial of service~\cite{alleaume2014,price2017}, reached through the quantum channel alone and, within one attempt, indistinguishable from a poorer channel (Sec.~\ref{sec:threat}); the pool-failure bound does not cover an adaptive adversary (Sec.~\ref{sec:threat}).

The margins are the real design lever (full sweep in the supplementary material). Scaling the phase and EC margins together at fixed $\ell=2077$ raises the key probability at $f=1\%$ from 0.862 to 0.967 at scale 1.5 and 0.977 at scale 2, but the budget grows to 493.7 and 736.0 million pulses and the undisturbed $\widehat D_{95}$ to 39.60 and 53.23~s. No tested scale both meets 35~s and stays above 0.95 at $f=1\%$; at 40~s, scale 1.5 does: a deadline implies a maximum tolerable disturbance.
\section{Discussion and Limitations}
\label{sec:limits}
The 1\% misalignment is optimistic: measured drone links report 2.5--3.4\%~\cite{B8_Drone-_and_Vehicle-Based_Quantum_Key_Distribution} and 2.2--6.9\%~\cite{tian}, where the window shrinks or vanishes (Sec.~\ref{sec:lossres}). Hover energy also bounds endurance: at 53--57~kJ per round and flight times of QKD-carrying drones from a few minutes~\cite{B8_Drone-_and_Vehicle-Based_Quantum_Key_Distribution} to over 40 minutes~\cite{tian}, a short-endurance battery supports only 4--9 rounds. If the pool is refilled at each landing rather than kept across battery changes, the floor does not bind and a smaller block (down to rule B1) suffices.

Security remains conditional on the inherited proof~\cite{a11}, successful authentication, and the error-correction proxy; neither a decoder nor the specified MAC is benchmarked. Measured reconciliation~\cite{mao} and privacy-amplification~\cite{tang} throughputs cannot be equated with a common workload rate; at tenfold cost these stages would add about 2.3~s. Confidence bounds address sampling error only, and no published scheme is reproduced as a comparator. Future work includes block redesign, a key reserve, flight between stations, and a variable-length protocol that shortens the key instead of aborting, with its own finite-key analysis. The supplementary material includes the analysis scripts and outputs for the reported results; some required inputs are not included.

\section{Conclusion}
Finite-key readiness, not computation or radio, dominates the latency, and hence the hover energy, of UAV edge computing with fresh keys: at 26~dB with benchmarked hashing it adds about 18~s per round (22~s in the high-cost stress reference). The pulse budget must lie between authentication-key replenishment, where that policy applies, and the deadline. For a fixed source, channel, and acceptance model, the authentication debit shifts the floor (281.8 to 422.7 million pulses for 0 to 2,880 bits per attempt); hashing cost does not. Conditional on the assumed acquisition probability, the selected configuration meets 35~s (held-out lower bound 97.3\%) at the least energy per task, and with benchmarked hashing a window opens even at 30~s. The largest tested loss with a non-empty 35-s window is 27~dB at 1\% misalignment and 22~dB at 2\% (held-out lower bound 96.9\% at 18--20~dB), none at 3\%; the stress reference lowers these limits by up to 2~dB at 10--20~Mworkbit/s. Small disturbances that leave $\varepsilon$ unchanged remove the service and drain authentication keys.
\FloatBarrier
\bibliographystyle{splncs04}
\bibliography{quav}
\FloatBarrier
\end{document}